\documentclass{WileyMSP-template}

\usepackage{siunitx}
\usepackage{wrapfig}
\usepackage{svg}
\usepackage{amsmath}
\usepackage{cite}

\begin{document}

\pagestyle{fancy}

\title{Growth windows of epitaxial $\textrm{Nb}_x\textrm{N}$ films on c-plane sapphire and their structural and superconducting properties}

\maketitle


\author{John G. Wright}
\author{Huili Grace Xing}
\author{Debdeep Jena}


\begin{affiliations}
John G. Wright*\\
Department of Materials Science and Engineering,\\
Cornell University\\
Ithaca, New York, 14853, USA\\
jgw92@cornell.edu\\
\hfill \break
Huili G. Xing\\
Department of Materials Science and Engineering,\\
Kavli Institute for Nanoscale Science,\\
\& School of Electrical and Computer Engineering,\\
Cornell University\\
Ithaca, New York, 14853, USA\\
\hfill \break
Debdeep Jena\\
Department of Materials Science and Engineering,\\
Kavli Institute for Nanoscale Science,\\
\& School of Electrical and Computer Engineering,\\
Cornell University\\
Ithaca, New York, 14853, USA\\

\end{affiliations}


\keywords{superconductor, heteroepitaxy, NbN, nitrides, molecular beam epitaxy}

\begin{abstract}

NbN films are grown on c-plane sapphire substrates by molecular beam epitaxy. The structural and superconducting properties of the film are characterized to demonstrate that growth parameters such as substrate temperature and active nitrogen flux effect the structural phase of films, and thereby the superconducting critical temperature. Four phases of NbN are identified for films grown in different conditions. In a novel finding, we demonstrate that atomically flat and highly crystalline $\beta$-$\textrm{Nb}_2\textrm{N}$ films can be grown at substrate temperatures of 1100 \degree C or higher, and that the superconducting critical temperature of phase pure $\beta$-$\textrm{Nb}_2\textrm{N}$ films is $0.35~K<T_c<0.6~K$, based on measurements of films grown at different substrate temperatures.

\end{abstract}


\section{Introduction}
The utility of epitaxial transition metal nitride (TMN) thin films has been demonstrated in a diverse array of applications, including the fabrication  of Josephson junctions \cite{Sun2016EpitaxialLayers,Makise2015FabricationApplications,Kaul2001InternallyApplications,Wang2013High-qualityDensity}, single photon detectors \cite{Miki2007NbNSubstrates,Cheng2020EpitaxialDetectors}, acoustic resonators \cite{Miller2020,Gokhale2020EpitaxialAcoustodynamics}, thermoelectric transducers \cite{Saha2011First-principlesConversion,Saha2018RocksaltMaterials}, epitaxial nucleation and sacrificial layers \cite{Katzer2015,Armitage2002Lattice-matchedSi,Meyer2016}, optical metamaterials \cite{Saha2014TiN/AlScNRange}, and the realization of topological electronic systems \cite{Dang2021AnSuperconductivity}. In addition to their intrinsic properties, a primary reason for the interest in TMN thin films is the ability to epitaxially integrate III-N semiconductors and TMNs to create structures which include semiconductor, metallic, superconductor, and ferroelectric thin films \cite{Jena2019TheEco-system}. As has been demonstrated, rock-salt cubic transition metal nitrides can be integrated with metastable rock-salt III-N semiconductors, enabling the growth of iso-structural metal-semiconductor heterostructures \cite{Rawat2009PseudomorphicSuperlattices,Saha2018RocksaltMaterials,Saha2011First-principlesConversion}. The inclusion of Sc in III-N semiconductors has been shown to increase the stability of the rock-salt phase, increasing the critical thickness of the metastable rock-salt semiconductor films \cite{Saha2014TiN/AlScNRange}.

Another method for epitaxial integration utilizes the growth of cubic TMNs oriented along the (1 1 1) body-diagonal with hexagonal III-N semiconductors grown along the c-axis \cite{Katzer2020,Yan2018,Dang2021AnSuperconductivity,Casamento2019MolecularAlN}. Although the crystal structures of the the two materials are markedly different in this case, the similarities between the structures and lattice parameters are sufficient to permit epitaxy, with the caveat that the growth of cubic films on hexagonal substrates results in multiple domains of the cubic film which differ in lattice orientation \cite{Wright2021UnexploredNbN,Casamento2019MolecularAlN,Kobayashi2020CoherentSputtering}.

This study focuses on the growth of NbN on sapphire by molecular beam epitaxy (MBE). We note that the low microwave dielectric loss of sapphire makes it an appropriate substrate material for noise-sensitive microwave devices, such as Josephson junction qubits. Despite numerous studies on the growth of NbN thin films by various methods over the past several decades \cite{Kobayashi2020CoherentSputtering,Wright2021UnexploredNbN,Hazra2016b,Kawaguchi1991PreparationMethod,Sowa2017,Oya1974,Ziegler2013,Kobayashi2020AutonomousSurfaces}, we present here novel findings which have implications for the growth of and applications of NbN films and NbN/III-N heterostructures. We demonstrate that the complexity of the Nb-N phase diagram has significant implications for the growth of NbN, as altering the growth conditions is shown to alter the phase, superconducting properties, surface roughness, metallic conductivity, and crystal quality of the films.

Of the results reported here, we would like to highlight the first unambiguous demonstration of epitaxial and single phase films of hexagonal $\beta$-$\textrm{Nb}_2\textrm{N}$. This result is important because $\beta$-$\textrm{Nb}_2\textrm{N}$ possesses excellent lattice and symmetry matching to wurtzite AlN. The cation atoms of $\beta$-$\textrm{Nb}_2\textrm{N}$ occupy a hexagonal close-packed lattice, identical to that of AlN, with a lattice parameter that differs by only 1.4\%\cite{Levinshtein2001PropertiesSiGe}. Therefore, $\beta$-$\textrm{Nb}_2\textrm{N}$ is an excellent material for heteroepitaxy with wurtzite III-N materials in general and AlN specifically, avoiding the multi-domain nature of cubic $\delta$-NbN films grown on hexagonal substrates \cite{Wright2021UnexploredNbN,Kobayashi2020AutonomousSurfaces}.

Contrary to previous reports that the superconducting critical temperature of $\beta$-$\textrm{Nb}_2\textrm{N}$ is between 8 and 12 K\cite{Zou2016,Gavaler1969}, we demonstrate that $\beta$-$\textrm{Nb}_2\textrm{N}$ is a superconductor with a critical temperature below 1 K. We have produced epitaxial $\beta$-$\textrm{Nb}_2\textrm{N}$ films on sapphire that are highly crystalline and atomically smooth, exhibiting a step and terrace surface morphology indicative of a step-flow growth mode. We discuss the discrepancies between our findings and previous reports on $\beta$-$\textrm{Nb}_2\textrm{N}$, and situate our findings in the broader context of studies of the Nb-N phase diagram.

\begin{figure}[h!]
  \centering
  \includegraphics[width=0.8\linewidth]{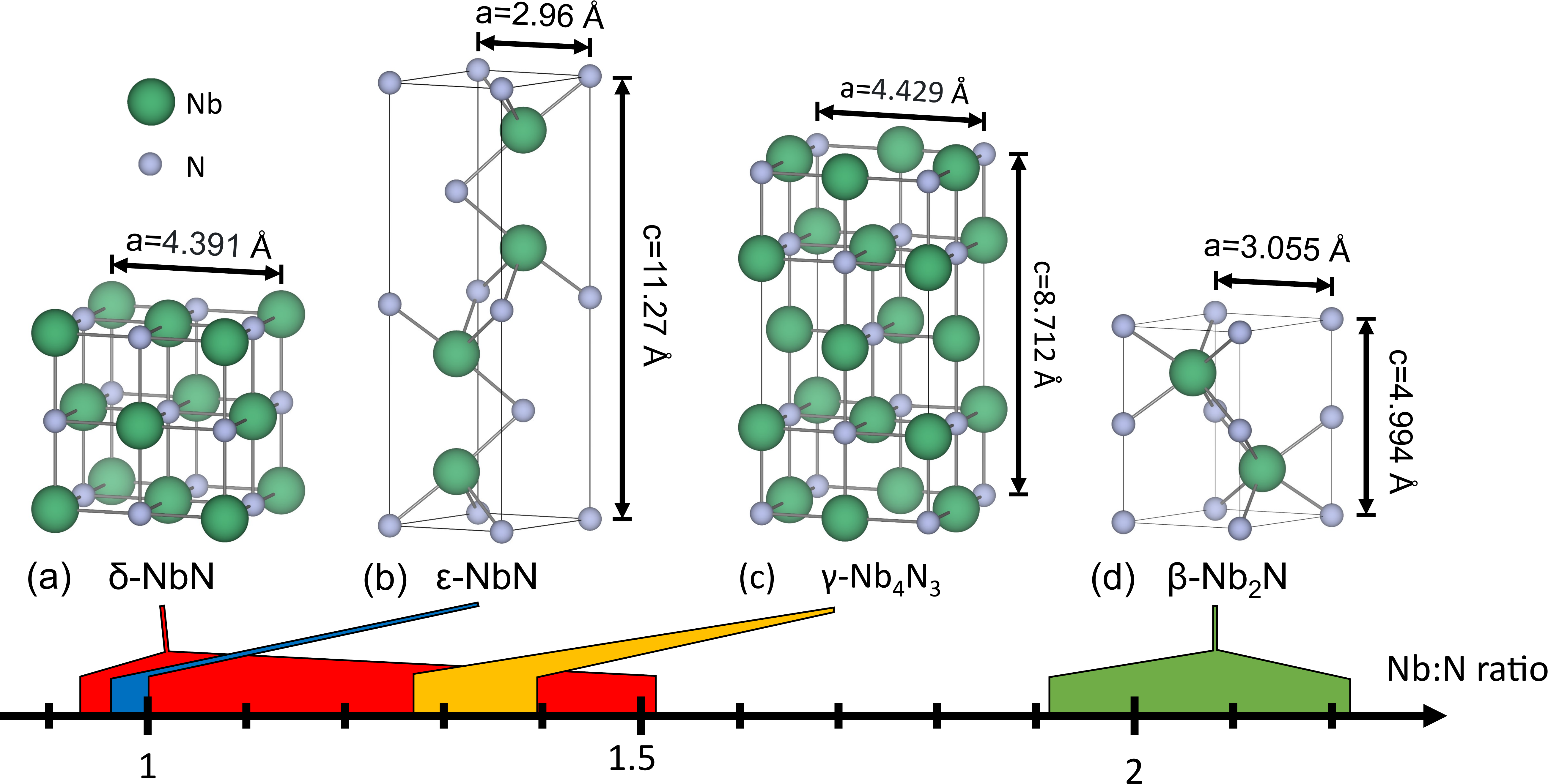}
  \caption{Unit cells, lattice parameters, and composition ranges of selected NbN crystal structures. Lattice parameters are obtained from (a) \cite{Lengauer1986PreparationDelta-NbNsub1-x} (b) \cite{Terao1965StructureNiobium} (c) \cite{Chervyakov1971NeutonN3} (d) \cite{Terao1965StructureNiobium}. Composition ranges are obtained from \cite{Brauer1964Drucksynthese-NbN,Joguet1998High-temperatureSystem,Lengauer2000Phase1400C}.}
  \label{fig:unit_cells}
\end{figure}

The canonical study of the Nb-N system by Brauer identified 5 phases of NbN, though one phase was determined to be metastable at all temperatures and compositions\cite{Brauer1960NitridesNiobium}. Brauer proposed a phase diagram, and later work using high-temperature XRD (HTXRD) by Lengauer slightly revised the NbN phase diagram \cite{Lengauer2000Phase1400C}. The 4 NbN phases included in Lengauer's phase diagram are shown in Figure 1. Like several other transition metal nitrides, such as ScN and TiN, NbN can adopt a B1 (rock salt) structure dubbed the $\delta$ phase as shown in Figure \ref{fig:unit_cells}(a). However, multiple experimental and theoretical studies have concluded that $\delta$-NbN is unstable at room temperature \cite{Brauer1960NitridesNiobium,Lengauer2000Phase1400C,Ivashchenko2010PhaseNitrides,Isaev2007PhononStudy}. The stable phase at room temperature and approximately 50\% nitrogen composition is the $\epsilon$ phase, a hexagonal TiP-type crystal shown in Figure \ref{fig:unit_cells}(b). $\gamma$-$\textrm{Nb}_4\textrm{N}_3$, shown in Figure \ref{fig:unit_cells}(c) is a stable tetragonal phase that is related to the $\delta$ phase by removal of half of the nitrogen atoms in alternating planes along the c-axis, accompanied by a slight distortion of the c-axis lattice parameter by about 2.8\%. Finally, $\beta$-$\textrm{Nb}_2\textrm{N}$ is a hexagonal phase with a NiAs-type crystal structure and approximately 50\% occupancy of the nitrogen sublattice, shown in Figure \ref{fig:unit_cells}(d). All NbN phases are metallic.

The structural differences between the various NbN phases are relevant for NbN thin film heteroepitaxy. Although both the $\beta$ and $\epsilon$ phases are hexagonal, they differ significantly in their lattice parameters, with the in-plane lattice parameter of the $\beta$ phase being 3.2\% larger than that of the $\epsilon$ phase, and the out-of-plane interplanar spacing of the $\epsilon$ phase being 12\% larger than that of the $\beta$ phase. On the other hand, despite the symmetry differences, the $\delta$ and $\beta$ structures both have close-packed cation sublattices with similar inter-cation spacing, with differences of 1.4\% and 2.0\% for the in-plane inter-cation spacing and the out-of-plane interplanar spacing respectively. Therefore, the $\delta$, $\gamma$ and $\beta$ phases all possess lattice parameters and structures which can enable heteroepitaxial growth with III-N semiconductor materials.

In Section \ref{sec:results} of this report, we present characterization of the electronic and structural properties of epitaxial NbN thin films grown by MBE on c-plane sapphire substrates. Section \ref{sec:growth_temp} examines the effect of the substrate growth temperature on the film properties. Section \ref{sec:nitrogen_flux} investigates the effects of nitrogen flux and the exposure to nitrogen post-growth on the NbN film properties. Section \ref{sec:discussion} summarizes and discusses the implications of our findings. Section \ref{sec:methods} presents details on the film growth and measurements.

\section{Results} \label{sec:results}
\subsection{Effects of growth temperature on the structural phase and superconducting properties of NbN}\label{sec:growth_temp}

\begin{figure}[h!]
  \centering
  \includegraphics[width=\linewidth]{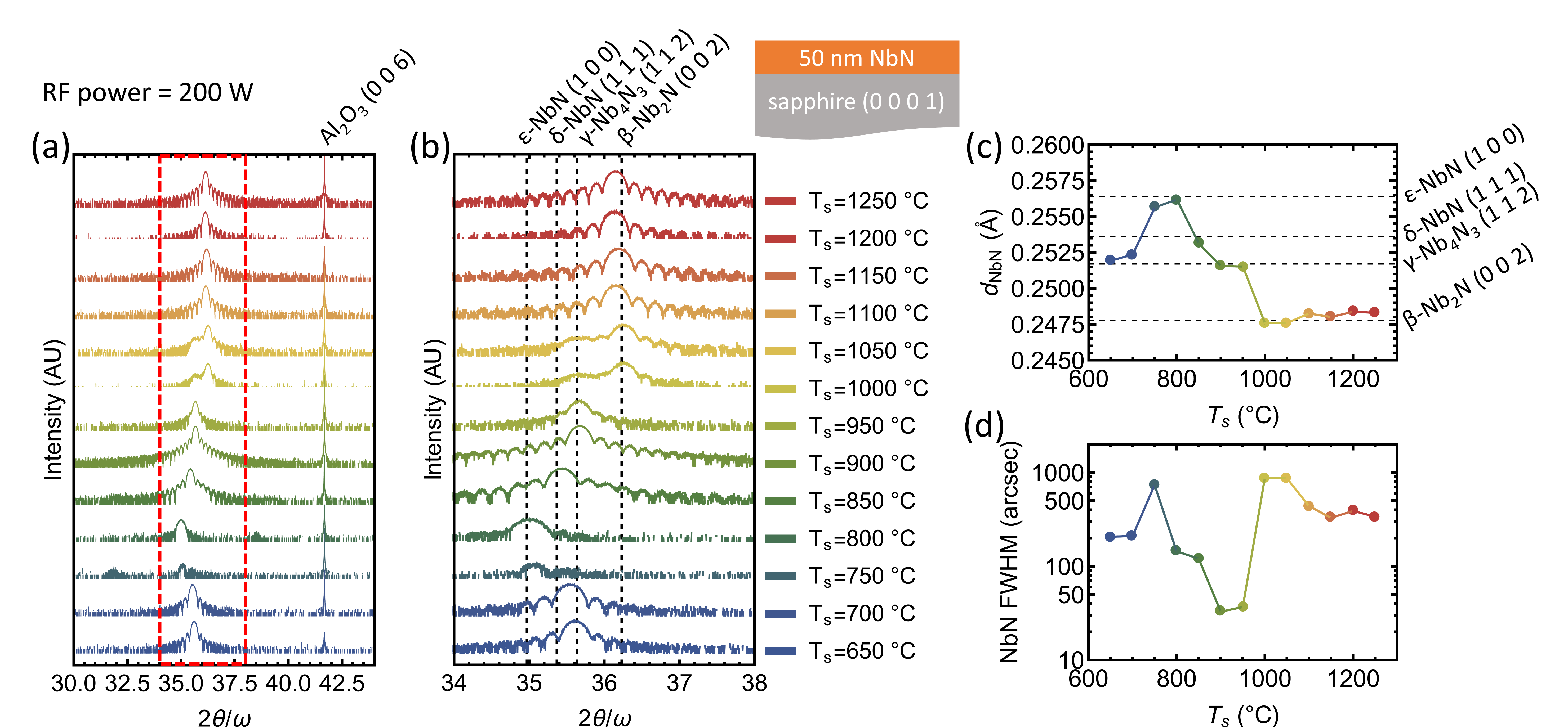}
  \caption{(a) XRD symmetric coupled scan for 50 nm NbN films grown on c-plane sapphire. (b) The symmetric XRD reflection due to the NbN films. (c) The symmetric interplanar spacing for NbN films as a function of growth temperature, with selected lattice planes of NbN displayed as dashed lines. (d) FWHM of the $\omega$ rocking curve of the symmetric NbN reflection.}
  \label{fig:XRD_temp}
\end{figure}

Figure \ref{fig:XRD_temp} shows symmetric high resolution X-ray diffraction (XRD) coupled scans of a series of approximately 50 nm NbN films grown on c-plane sapphire substrates. The films are grown with identical nitrogen and niobium fluxes, but different substrate temperatures. Figure 2(a) shows the NbN peaks nearest to the sapphire (0 0 6) reflection, which is observed at $2 \theta \cong 41.7^\circ$. The substrate temperature is varied between between 650 $^\circ \textrm{C}$ and 1250 $^\circ \textrm{C}$ in 50 $^\circ \textrm{C}$ intervals for different growths. Although all XRD measurements reveal a reflection from the NbN films in the range $30^\circ<2 \theta<40^\circ$, the peak location varies significantly as a function of substrate temperature. Figure \ref{fig:XRD_temp}(b) shows the same coupled scans, but focused on the reflection due only to the NbN films. We observe that for certain substrate temperatures, such as 650 $^\circ \textrm{C}$, 900 $^\circ \textrm{C}$, and 1200 $^\circ \textrm{C}$, the films exhibit a single reflection and clear Pendell$\textrm{\"o}$sung fringes, evidence that the films are single phase with smooth interfaces with the sapphire substrate and smooth surfaces. However, at other substrate temperatures, such as 750 $^\circ \textrm{C}$, and 1000 $^\circ \textrm{C}$, weak peaks or multiple peaks, and a lack of Pendell$\textrm{\"o}$sung fringes are observed. We can therefore immediately conclude that substrate temperature has a profound effect on the crystallinity and surface morphology of the NbN films grown by MBE on c-plane sapphire. 

Figure \ref{fig:XRD_temp}(c) shows the out-of-plane interplanar spacing $d_{NbN}$ extracted from the symmetric XRD coupled scans for each NbN film as a function of growth temperature. In the case that two peaks are evident in the XRD scan, such as for the film grown at $T_s=1000~^\circ{}\textrm{C}$, only the most intense peak is represented. The expected interplanar spacing of specific planes and phases of NbN are represented as dashed lines on the graph. It is worth mentioning that we feel correspondences between interplanar spacings of specific phases, and measured values from XRD on a graph such as this should not be viewed as a method of definitive phase identification for a material with so many phases such as NbN. Firstly, several phases, such as the tetragonal $\textrm{Nb}_4 \textrm{N}_5$\cite{Terao1971NewNitride}, the primitive cubic NbN\cite{Treece1995a}, and the hexagonal $\delta$'-$\textrm{NbN}$ \cite{Brauer1960NitridesNiobium}, are left off of this graph to enhance readability. Secondly, many of the interplanar distances are quite similar, and strain and compositional variation can lead to changes in interplanar spacing within a phase, making phase identification from a single interplanar spacing rather unreliable.

However the preliminary conclusion evident in Figure \ref{fig:XRD_temp}(c), which we substantiate further later, is that changes in growth temperature lead to changes in the NbN structural phase. In different films, we observe all four NbN phases displayed in Figure \ref{fig:unit_cells}: the hexagonal $\epsilon$ phase, the cubic $\delta$ phase, the tetragonal $\gamma$ phase, and the hexagonal $\beta$ phase. Table \ref{tab:table1} shows a summary of the phase of each film as a function of growth temperature.  We note that changes in lattice parameter as a function of growth temperature highly similar to what we observe in Figure \ref{fig:XRD_temp}(c) are reported by Kobayashi et al. in sputtered NbN films grown on AlN \cite{Kobayashi2020CoherentSputtering}, however, these changes are attributed by the authors to strain effects. We conclusively demonstrate that the changes we observe in film properties as a function of growth temperature are a result of changes in the \emph{structural phase} of the films. We therefore believe our conclusions on the changes in NbN thin films as a function of growth temperature have general implications for the growth of NbN by a variety of thin film growth methods.

One of the primary results of this report is the finding that highly crystalline, phase pure epitaxial $\beta-\textrm{Nb}_2\textrm{N}$ films can be grown on sapphire substrate by MBE at sufficiently high substrate temperature. In our experiment we determine that all films grown at 1100 $\degree$C and above are purely the $\beta$ phase.

\begin{table}[h!]
 \centering
 \begin{tabular}{m{2cm} m{3.5cm} m{2.5cm} m{2.5cm} m{5cm}}
    \hline\hline
    $T_s$ ($\degree$C) & Phase(s) & $T_c$ (K) & Orientation to $\textrm{Al}_2 \textrm{0}_3$ (0 0 0 6) & Orientation to $\textrm{Al}_2 \textrm{0}_3$ (1 1 $\bar{2}$ 0)\\
    \hline
    \\[0pt]
    650 - 700 & $\delta$-NbN  & 13 K - 14 K & (1 1 1) & (1 1 $\bar{2}$) \& ($\bar{1}$ $\bar{1}$ 2) \\
    750 - 800 & $\epsilon$-NbN  & NA  & (1 0 $\bar{1}$ 0) & (0 0 0 1) \\
    850 & $\delta$-NbN $\&$ $\epsilon$-NbN  & 15.0 K & - & - \\
    900 & $\gamma$-$\textrm{Nb}_4\textrm{N}_3$  & 10.8 K & (1 1 2) & (1 1 $\bar{4}$), ($\bar{1}$ $\bar{1}$ 4), ($\bar{2}$ 1 2), (2 $\bar{1}$ $\bar{2}$), (1 $\bar{2}$ 2), \& ($\bar{1}$ 2 $\bar{2}$) \\
    950 - 1050 & $\gamma$-$\textrm{Nb}_4\textrm{N}_3~\&~\beta$-$\textrm{Nb}_2\textrm{N}$  & 7 K - 5 K & - & - \\
    1100 - 1250  & $\beta$-$\textrm{Nb}_2\textrm{N}$  & 0.4 K - 0.6 K & (0 0 0 2) & (1 0 $\bar{1}$ 0) \\
    \hline\hline
  \end{tabular}
  \caption{The structural phase and superconducting transition temperature of NbN films grown at different substrate temperatures.}
  \label{tab:table1}
\end{table}

\begin{figure}[h!]
  \centering
  \includegraphics[width=\linewidth]{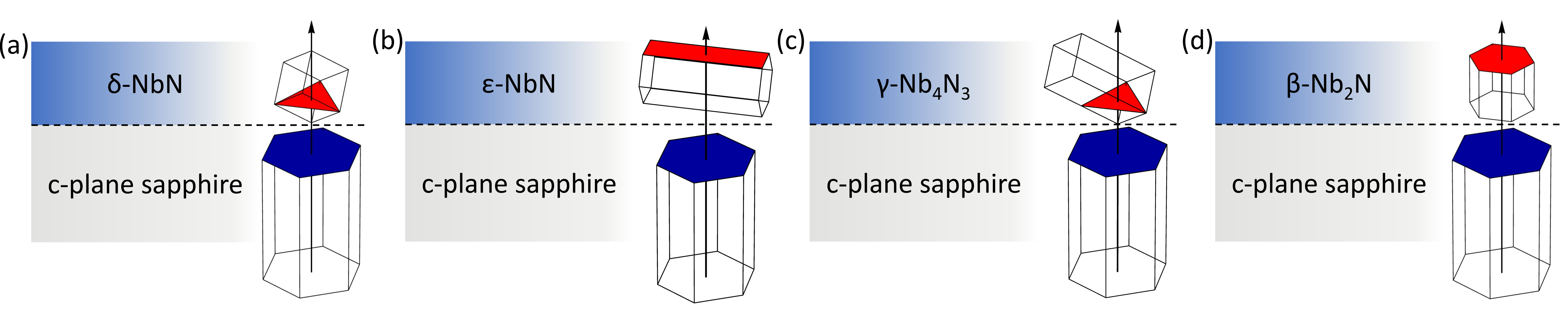}
  \caption{The experimentally determined orientations of various NbN unit cells on c-plane sapphire: (a) $\delta$-NbN, (b) $\epsilon$-NbN, (c) $\gamma$-$\textrm{Nb}_4\textrm{N}_3$, (d) $\beta$-$\text{Nb}_2\textrm{N}$
  }
  \label{fig:NbN_unit_cell_orientation}
\end{figure}

Figure \ref{fig:XRD_temp}(d) shows the measured symmetric $\omega$ rocking curve full-width at half maximum (FWHM) as a function of substrate temperature for the same NbN films shown in Figure \ref{fig:XRD_temp}(a)-(c). The $\omega$-scan FWHM assesses broadening due to lattice tilt, the finite size of crystallites, and the presence of dislocations. We observe that several of the films which exhibit weak peaks or multiple peaks in the coupled XRD scans, such as the films grown at 750 $\degree$C or 1000 $\degree$C, show large values of the FWHM. The minimum observed value for the symmetric rocking curve FWHM is 33", a value which indicates the degree of epitaxial orientation of the NbN is very high.

All films exhibit out-of-plane epitaxial crystallographic orientations in the symmetric XRD measurements, and therefore conclusive phase characterization by XRD is more challenging than would be the case for randomly oriented films. Our method of definitive phase characterization relies on using the symmetric XRD and RHEED measurements to form a hypothesis for the phase and orientation of the film, and then using XRD reciprocal space mapping to detect the presence or absence of an asymmetric XRD peak in the location expected for a given phase and orientation. Care must be taken to ensure that the chosen peaks conclusively establish the presence of a given phase, as the structural similarities between the different phases of NbN often leads to reciprocal lattice points that would overlap.

\begin{figure}[h!]
  \centering
  \includegraphics[width=1\linewidth]{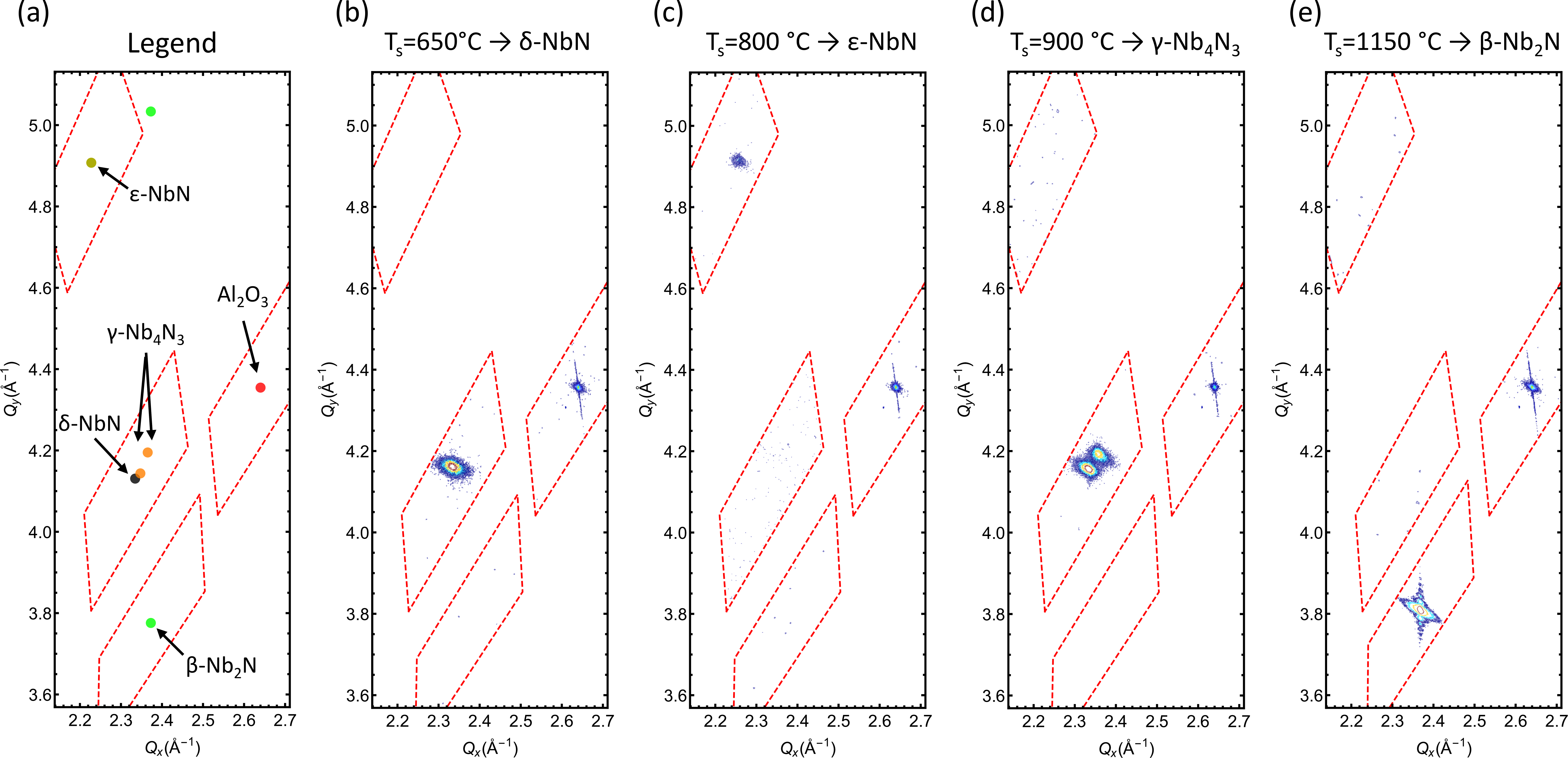}
  \caption{XRD-RSM of NbN films on c-plane sapphire with different structural phases grown at different temepratures. (a) A legend showing the expected locations of reciprocal lattice points for a variety of NbN phases as well as the sapphire substrate. (b) $\delta$-NbN grown at 650 $\degree$C. (c) $\epsilon$-NbN grown at 800 $\degree$C, (d) $\gamma$-$\textrm{Nb}_4\textrm{N}_3$ grown at 900 $\degree$C. (e) $\beta$-$\textrm{Nb}_2\textrm{N}$ grown at 1150 $\degree$C. All measurements are taken along the sapphire (1 $\bar{1}$ 0) zone axis, The sapphire peak shown in each scan is the (1 1 $\bar{2}$ 9) reflection.}
  \label{fig:RSM}
\end{figure}

Figure \ref{fig:RSM} shows the results of the reciprocal space mapping for 4 representative NbN films grown at different temperatures. All regions of reciprocal space outlined in red were mapped for each film. The same zone axis was used for each film, and the orientation and alignment of the measurement was confirmed by measuring the presence of the sapphire (1 1 9) reflection for each film. We see that different reciprocal lattice points are detected for each film, allowing us to conclude the structural phase and orientation of each film. The phase and orientation determinations from RSM match the expected orientations determined by the out-of-plane lattice parameters shown in Figure \ref{fig:XRD_temp}(c), and the orientations of each phase are summarized in Table \ref{tab:table1} and Figure \ref{fig:NbN_unit_cell_orientation}.

We conclude that, at the lowest growth temperatures, the films adopt the cubic $\delta$ structure. The orientation of the $\delta$-NbN unit cell relative to the sapphire substrate is shown in Figure \ref{fig:NbN_unit_cell_orientation}(a). Two orientations of $\delta$-NbN, related by a 180$\degree$ rotation about the growth axis, are symmetrically identical with respect to the c-plane sapphire substrate, and both orientations are present in the films. Films grown at temperatures of 750 $\degree$C and 800 $\degree$C possess the hexagonal $\epsilon$ phase. We note that the c-axis of the $\epsilon$ phase is not aligned to the sapphire c-axis, as shown in Figure \ref{fig:NbN_unit_cell_orientation}.

Further increase of the substrate temperature causes a transition to the tetragonally-distorted $\gamma$ phase. The structural similarities between the $\gamma$ and $\delta$ phases can make them difficult to distinguish. However, as can be observed in Figure \ref{fig:RSM}(d), in the location of a single reciprocal lattice point that is expected for the $\delta$ phase, two reciprocal lattice points are expected for the $\gamma$ phase as a result of the small tetragonal distortion of one of the cubic lattice vectors. If the tetragonal crystal possessed a single orientation, these two peaks would not lie in the same plane in reciprocal space. However, as can be seen in Figure \ref{fig:NbN_unit_cell_orientation}, 6 different orientations of the tetragonal unit cell, related by rotation by 60$\degree$ about the out of plane vector, are symmetrically identical with respect to the underlying sapphire crystal.

Finally, films grown at the highest growth temperatures possess the hexagonal $\beta$-$\textrm{Nb}_2\textrm{N}$ phase, with the c-axis of the film aligned to that of the sapphire substrate. The unit cell of the $\beta$-$\textrm{Nb}_2\textrm{N}$ adopts the same in-plane orientation as AlN and GaN grown on sapphire \cite{Ruterana2003NitrideDevices}, which is not supprising given the similarities between the AlN and $\beta$-$\textrm{Nb}_2\textrm{N}$ lattices. From our XRD measurements, we determine that the lattice parameters of the $\beta$-$\textrm{Nb}_2\textrm{N}$ film are $a=3.066 \pm 0.004 \si{\angstrom}$ and $c=4.965 \pm 0.001 \si{\angstrom}$.

\begin{figure}[h!]
  \centering
  \includegraphics[width=0.9\linewidth]{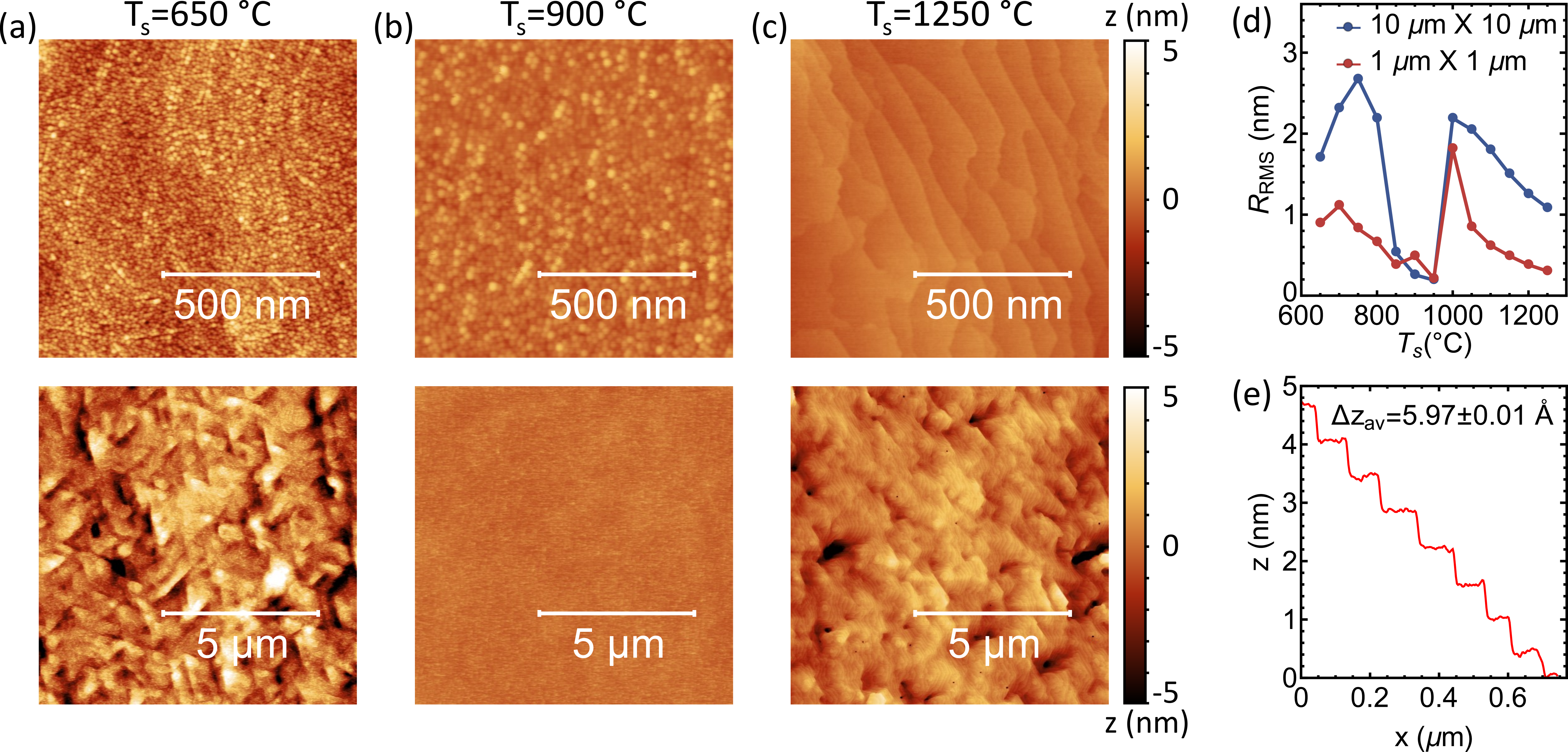}
  \caption{AFM surface height maps for NbN films grown on sapphire at (a) $T_s=650 ~\degree \textrm{C}$, (b) $T_s=900~\degree \textrm{C}$, (c) $T_s=1250 ~\degree \textrm{C}$. (d) RMS roughness of NbN films as a function of $T_s$. (e) Surface height profile of atomic steps of NbN film grown at $T_s=1250~\degree \textrm{C}$.}
  \label{fig:AFM}
\end{figure}

Analysis of the surface morphology of the NbN films reveals significant variation in growth mode as a function of substrate temperature, as shown in Figure \ref{fig:AFM}(a)-(c). We highlight the  observation that, for $T_s$ greater than approximately 1100 $\degree$C, the film surface morphology shows a clear step-and-terrace morphology, evidence of a step-flow growth mode. An example of such morphology is seen in Figure \ref{fig:AFM}(c). This is, to our knowledge, the first report of a step-flow growth mode for NbN films of any phase. All films in our study observed to possess the step-and-terrace surface morphology are $\beta$-$\textrm{Nb}_2\textrm{N}$ phase. Figure \ref{fig:AFM}(a)-(b) show that, at lower values of $T_s$, the features are consistent with a Volmer-Weber, or island formation, growth mode. 

Figure \ref{fig:AFM}(d) shows the variation of the root mean square roughness of the films. The $\beta$-$\textrm{Nb}_2\textrm{N}$ films ($T_s>1050~\degree$C) show decreasing roughness as $T_s$ is increased. Figure \ref{fig:AFM}(e) shows the height profile of a line scan across the surface of the film shown in Figure \ref{fig:AFM}(c). The average step-height, $\Delta z_{av}$, is measured to be 5.97 $\si{\angstrom}$. From the XRD measurement in Figure \ref{fig:RSM}(c)\&(d), the measured out of plane interplanar spacing for the same film is 2.48 $\si{\angstrom}$. There is therefore poor correspondence between any integer multiple of the out-of-plane lattice parameter and the height of the atomic steps. We hypothesize that the steps are likely bilayer steps, but that oxidation of the surface upon exposure to atmosphere alters the crystal structure of the surface layers and thereby alters the height of the steps.

\begin{figure}[h!]
  \centering
  \includegraphics[width=0.8\linewidth]{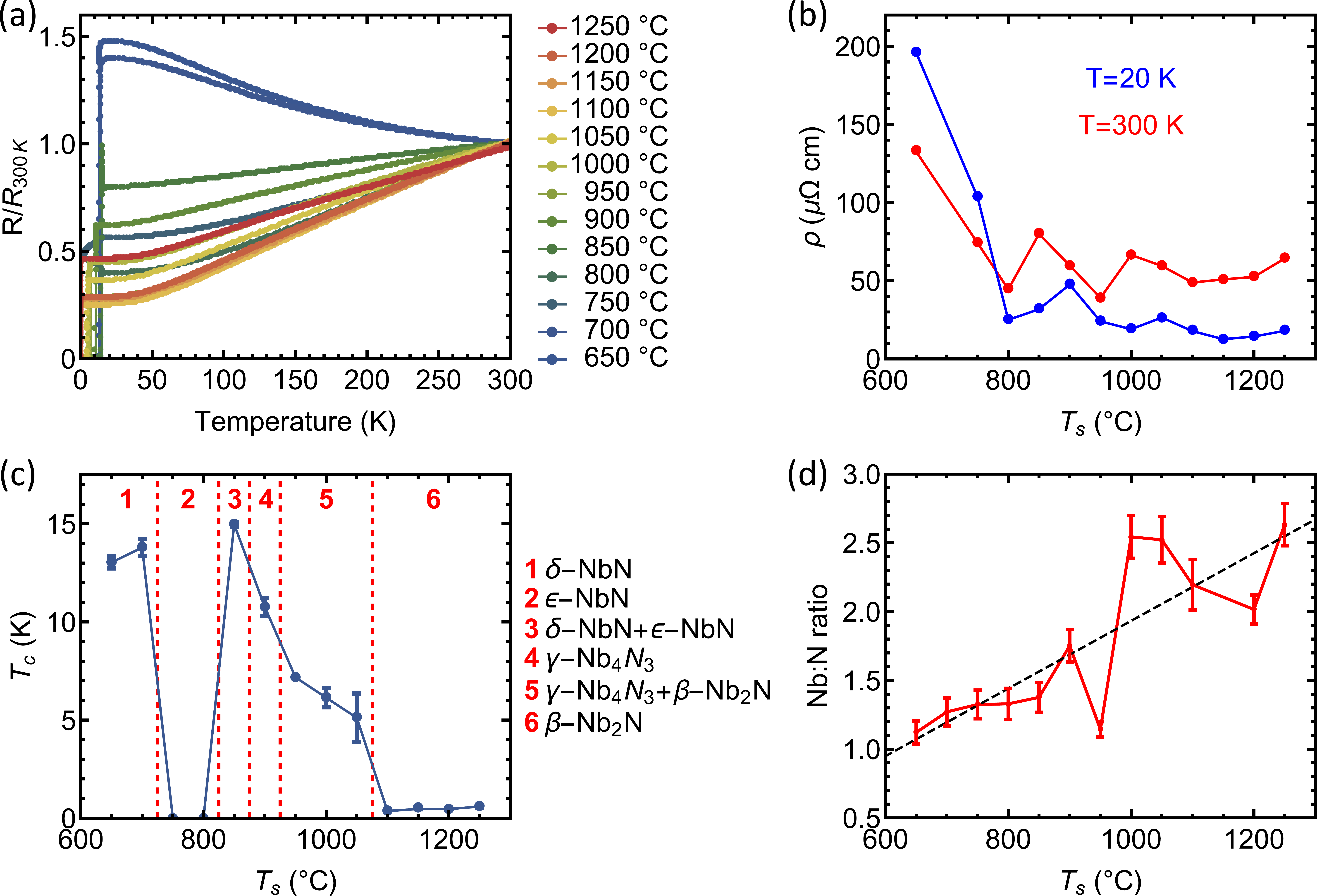}
  \caption{(a) Normalized resistance of NbN films as a function of temperature for films grown at different values of $T_s$. (b) Resistivity at 300 K for NbN films as a function of $T_s$. (c) $T_c$ of NbN films as a function of $T_s$. (d) Nb:N atomic ratio measured by EDS for films grown at different $T_s$.}
  \label{fig:R_vs_T}
\end{figure}

Figure \ref{fig:R_vs_T}(a) shows the measured film resistances as a function of temperature, scaled to the resistance at 300 K. We observe a general trend that films grown at higher substrate temperature show more metallic type conduction, with a positive temperature coefficient of resistivity. The two films grown at the lowest $T_s$ show negative temperature coefficient of resistivity. The resistivity of films at both 20 K and 300 K as a function of substrate temperature is displayed in Figure \ref{fig:R_vs_T}(b). All values of resistivity are within the range typically observed for NbN thin films\cite{Pan1983InvestigationFilms,Shoji1992,Chockalingam2008SuperconductingFilms}. As a result of the greater residual-resistivity ratio (RRR) for films grown at high temperatures, we see that the resistivity at 20 K shows a general trend of decreasing resistivity at 20 K as substrate temperature increases.

The superconducting critical temperature ($T_c$) shows a strong dependence on $T_s$, as shown in Figure 5(c). The observed changes in structural phase of the films as $T_s$ is varied provide insight into the seemingly erratic variations of $T_c$. We see that the highest $T_c$ is observed in the three films which contain the $\delta$-NbN phase ($T_s=650~\degree \textrm{C},700~\degree \textrm{C}, 850~\degree$C).

Our results show no evidence of superconductivity for films which possess the $\epsilon$-NbN structure, down to the minimum measurement temperature of 0.4 K. Previous studies have reported significantly different values for the transition tempearture of $\epsilon$-NbN. Oya et al. \cite{Oya1974}, who fabricated NbN films using a vapor phase growth with $\textrm{NbCl}_5$ and $\textrm{NH}_3$ precursors, concluded based on measurements of films containing both polycrystalline $\delta$-NbN and $\epsilon$-NbN that $\epsilon$-NbN exhibited no superconductivity above 1.77 K, the minimum measurement temperature used in the study.

A more recent report by Zou et al. reported a transition at 11.6 K for polycrystalline $\epsilon$-NbN material fabricated by annealing NbN starting material at high temperature and pressure \cite{Zou2016}, however in this study they also showed evidence of some $\delta$-NbN material in the $\epsilon$-NbN samples using XRD analysis. The films showed evidence of a two-step transition to the superconducting state, both in resistance and magnetic moment measurements, with the first step occurring near 17.5 K, consistent with the presence of $\delta$-NbN in the samples, which is the only NbN phase with a $T_c$ as high as 17 K.

We therefore hypothesize that reports of a higher critical temperature for $\epsilon$-NbN are due to material which contain both $\epsilon$-NbN and $\delta$-NbN, and that the observed superconductivity is due entirely to the presence of the $\delta$-NbN and the superconducting proximity effect. Our results support Oya's initial finding, and lead us to conclude that $\epsilon$-NbN exhibits no superconductivity down to at least 0.4 K.

The critical temperature we measure for the film with $\gamma$-$\textrm{Nb}_4\textrm{N}_3$ structure, $T_c=10.8~\textrm{K}$, is consistent with previous findings from Oya \cite{Oya1974}, who also reports a $T_c$  $\sim 11~\textrm{K}$ for $\gamma$-$\textrm{Nb}_4\textrm{N}_3$.

For films grown at substrate temperatures between the values that lead to growth of pure $\gamma$ ($T_s=900~\degree$C) and pure $\beta$ ($T_s=1100~\degree$C) films, we see a gradual decrease in $T_c$ from the value for pure $\gamma$ phase films as $T_s$ is increased, as seen in Figure \ref{fig:R_vs_T}(c). We hypothesize that the decreased $T_c$ can be attributed to the proximity effect between $\gamma$ and $\beta$ phase regions. The disappearance of the $\gamma$ phase and the appearance of pure $\beta$ phase films at $T_s=1100~\degree$C is accompanied by a drop in $T_c$ to approximately 0.36 K. All films which exhibit pure $\beta$ phase structure exhibit $0.35~\textrm{K}<T_c<0.6~\textrm{K}$.

Several previous studies on the growth and superconducting properties of $\beta$-$\textrm{Nb}_2\textrm{N}$ report conflicting results, and none that we are aware of are consistent with our findings. Skokan et al. report the growth of $\beta$ phase NbN thin films by sputtering, but the results of their phase characterization by XRD are consistent with films of mixed $\alpha$-Nb/$\beta$-$\textrm{Nb}_2\textrm{N}$ structure \cite{Skokan1980SuperconductingImplantation}. Their measurements of $T_c$ are also consistent with such a mixed phase material, showing initial drops in resistance at around 9 K (near the value of $T_c$ for Nb), and a broad transition that reaches zero only at 2 K. Gavaler et al. report films that exhibit XRD evidence of the $\beta$ phase, but also mention that these films exhibit evidence of additional peaks in XRD that could not be indexed. No additional information is given about the sample, and the XRD data are not presented \cite{Gavaler1969}. For these films they report a $T_c$ of 8.6 K.

Katzer and Nepal et al. report the growth of $\beta$-$\textrm{Nb}_2\textrm{N}$ films on 6H-SiC substrate by MBE \cite{Katzer2015,Nepal2016CharacterizationSubstrates} at temperatures at and around $800~\degree \textrm{C}$, and report a value of value of $T_c=11.5~\textrm{K}$ \cite{Yang2017MagneticSubstrate}. For phase characterization, they utilize both symmetric and asymmetric XRD. However, when the expected epitaxial orientation of each phase on 6H-SiC substrate is considered, we find that each of the asymmetric $\beta$ phase peaks they utilized for asymmetric XRD characterization is nearly coincident in reciprocal space with the expected location of a $\delta$ phase or $\gamma$ phase peak. Therefore, the asymmetric XRD used was unlikely to be able to distinguish between $\beta$ and $\delta$ phase films. Interestingly, Nepal et al. actually show double peaks in their asymmetric XRD characterization of their NbN films, consistent with the expected double peaks of $\gamma$ phase films such as is shown in Figure \ref{fig:RSM}(d) \cite{Nepal2016CharacterizationSubstrates}. Furthermore, the surface of their films shows the parallel domain boundaries characteristic of cubic transition metal nitride films grown on hexagonal substrates, such as reported by Kobayashi, Casamento, and Wright \cite{Kobayashi2020AutonomousSurfaces,Casamento2019MolecularAlN,Wright2021UnexploredNbN}. The explanation that their films actually possess the $\gamma$ phase and not the $\beta$ phase is consistent with their observation of $T_c=11.5~\textrm{K}$, a value consistent with the $T_c$ of $\gamma$ phase films in this study and those reported by Oya \cite{Oya1974}. Finally, Gajar et al. reported the growth of mixed phase films of $\beta$-$\textrm{Nb}_2\textrm{N}$ and $\textrm{Nb}_4\textrm{Nb}_5$, which exhibit broad superconducting transitions around 1 K \cite{Gajar2019SubstrateStudy}.

We therefore believe that all previous reports of the critical temperature of $\beta$-$\textrm{Nb}_2\textrm{N}$ suffer from either phase-impure material, or from mistaken phase characterization. We thus conclude that, contrary to previous reports, $\beta$-$\textrm{Nb}_2\textrm{N}$ is a superconductor with a critical temperature below 1 K. The maximum critical temperature we observe for $\beta$-$\textrm{Nb}_2\textrm{N}$ is 0.59 K. For this film, we observe a single transition directly to a zero resistance state that is sharp, with a transition width given by $\Delta T_c=0.05 \textrm{K}$.

Our observation that the phase of the MBE grown NbN is strongly determined by the growth temperature is, we believe, not a function of the intrinsic temperature stability of the different NbN phases. Previous HTXRD studies have shown that the $\gamma$ phase is stable up to temperatures between 1100 \degree C and 1200 \degree C, after which it converts into nitrogen deficient $\delta$-NbN \cite{Berger1997TheTransition}. $\epsilon$-NbN is stable up to a temperature of approximately 1300 \degree C, above which it too converts into the $\delta$ phase\cite{Lengauer2000Phase1400C}. The $\beta$ and $\delta$ phases are expected to be stable to much higher temperatures\cite{Brauer1960NitridesNiobium}, while $\delta$-NbN is thermodynamically unstable below between 1070 \degree C and 1300 \degree C, depending on composition\cite{Berger1997TheTransition}.

Therefore, the variations in phase we observe as a function of growth temperature do not intuitively follow from the Nb-N phase diagram. Instead, we propose that the phase variations are best understood as resulting from changes in the \emph{nitrogen content} of the films as a function of growth temperature. Figure 5(d) shows the measured Nb:N ratio of the films measured by scanning electron microscope (SEM) based energy dispersive X-ray spectroscopy (EDX), which we calculate using only the measured quantity of Nb and N. Given the lack of a standard sample used for calibration, the inherent challenges in detection of characteristic X-rays from nitrogen, and other challenges in accurate compositional determination for such thin films by EDX, we view the measured variation in composition as meaningful, but the absolute values of composition as imprecise. The measured composition as a function of growth temperature shows an overall trend towards higher Nb:N ratio as the substrate growth temperature is increased. The $\epsilon$ phase has the narrowest composition window in the NbN phase diagram, and several experimental studies have shown that the composition of $\epsilon$-NbN lies within a range of 49\% and 52\% nitrogen \cite{Brauer1960NitridesNiobium}\cite{Lengauer2000Phase1400C}. Our measured values of composition for our $\epsilon$ phase films are closer to 43\%, and we therefore suspect that our EDS measurement of the composition may be systematically under-valuing the nitrogen composition.

\subsection{Effects of nitrogen flux and post growth nitrogen conditions on the structural phase and superconducting properties of NbN}\label{sec:nitrogen_flux}

There are several factors that could potentially contribute to the observed variation of nitrogen concentration in the films as a function of substrate temperature. Variation in the free energy of nitrogen vacancy formation, variation in the sticking coefficient of nitrogen, and variation in the equilibrium partial pressure of nitrogen over NbN films could all contribute to changes in the film composition with temperature.

\begin{figure}[h!]
  \centering
  \includegraphics[width=0.7\linewidth]{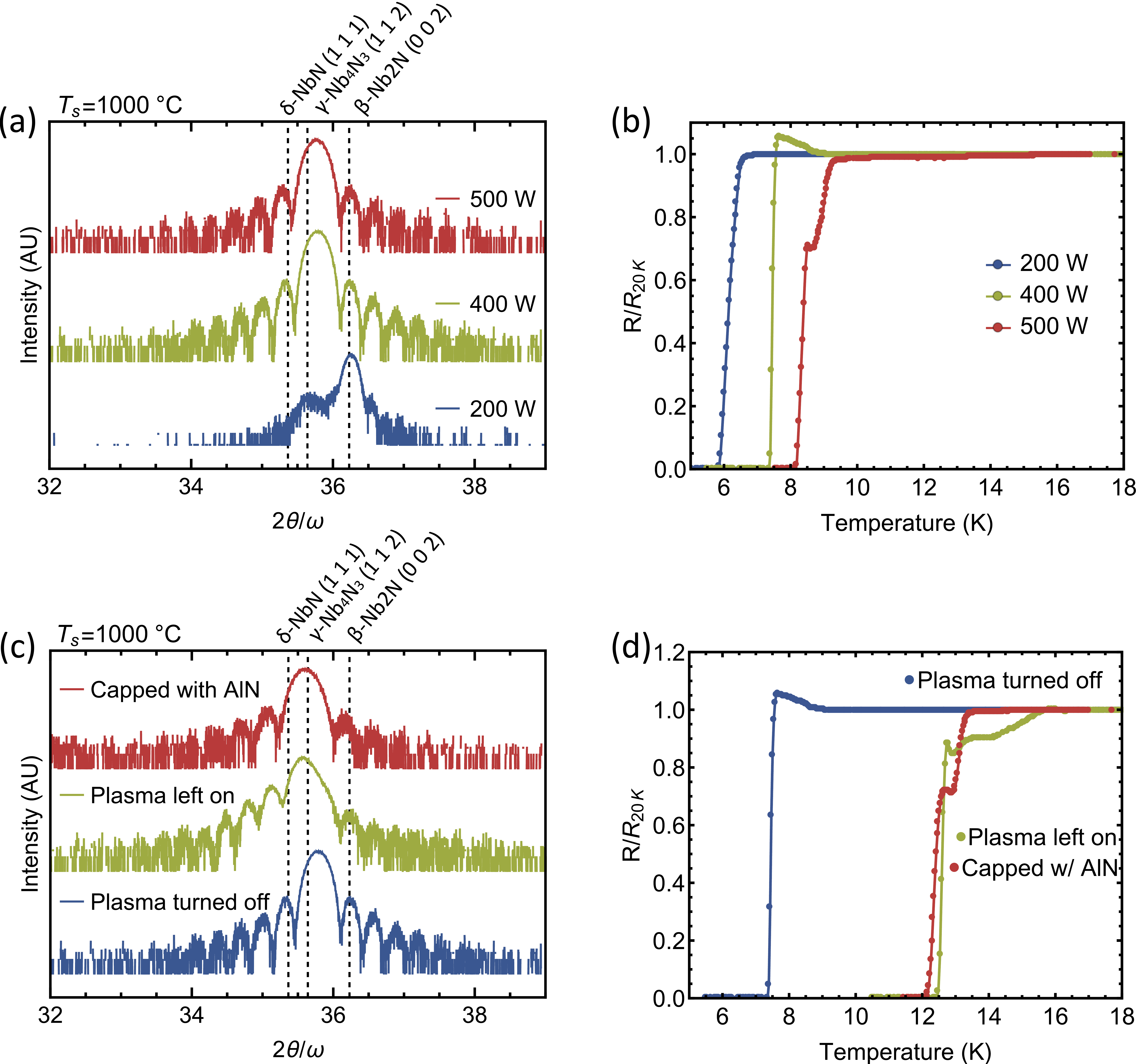}
  \caption{(a) XRD coupled scans and (b) \emph{R} vs \emph{T} measurements of NbN films grown at $T_s=1000~\degree$C and different nitrogen plasma conditions. Films are labelled by the RF power of the plasma source. (c) XRD coupled scans and (d) \emph{R} vs \emph{T} measurements of NbN films grown at $T_s=1000~\degree$C and 400 W nitrogen plasma conditions, but treated with different processes after growth.}
  \label{fig:Plasma_post_growth}
\end{figure}

We performed several experiments to understand the relationship between film composition, substrate temperature, structural phase, and nitrogen flux. In one experiment, we chose a substrate temperature (1000 \degree C) which had initially yielded a mixed $\gamma$ and $\beta$ phase film, as shown by the presence of two peaks in Figure \ref{fig:XRD_temp}(b). We grew three films at this same substrate temperature using progressively higher nitrogen fluxes, which we controlled by increasing both the nitrogen flow rate and the RF power for the nitrogen plasma source in the MBE system. Figure 6(a) shows symmetric XRD scans for these 3 films, with the nitrogen RF plasma power used to label the films. Higher RF plasma powers result in higher nitrogen flux. We observe that while the film grown with the lowest nitrogen flux (200 W) exhibits both $\gamma$-$\textrm{Nb}_4\textrm{N}_3$ and $\beta$-$\textrm{Nb}_2\textrm{N}$ phases, the films grown with the higher nitrogen fluxes (400 W and 500 W) are pure $\gamma$-$\textrm{Nb}_4\textrm{N}_3$ phase. We therefore conclude that, at a constant substrate temperature, a higher nitrogen flux can stabilize the more nitrogen rich of two NbN phases. Put another way, while increasing $T_s$ from 900 \degree C to 1000 \degree C leads to a transition from pure $\gamma$-$\textrm{Nb}_4\textrm{N}_3$ to mixed $\gamma$-$\textrm{Nb}_4\textrm{N}_3$/$\beta$-$\textrm{Nb}_2\textrm{N}$, increasing the nitrogen flux at 1000 \degree C leads to a transition back to pure $\gamma$-$\textrm{Nb}_4\textrm{N}_3$, and in this sense temperature and nitrogen flux play opposing roles in regulating the nitrogen content of the films, and thereby controlling the structural phase. The structural change from a mixed $\gamma$ and $\beta$ phase for 200 W plasma conditions to pure $\gamma$ phase for 400 W and 500 W conditions is also reflected in the increase in the superconducting critical temperature of the films, as shown in Figure \ref{fig:Plasma_post_growth}(b). 

Previous reports indicate that the nitrogen diffusion in NbN becomes appreciable at temperatures above 1000 \degree C \cite{Musenich1994GrowthTemperature}. In fact, based on the nitrogen diffusion coefficient in NbN reported by Musenich et al., and the limited rate at which we can cool the substrate material after growth ($\sim$30 \degree C/minute), we conclude that for all temperatures above approximately 900 \degree C, the cumulative nitrogen diffusion length after growth is greater than the film thickness ($\sim$50 nm). We therefore hypothesized that, for substrate growth temperatures above around $T_s>900~\degree$C, that nitrogen would be able to diffuse through the film and desorb after growth, leading to changes in film composition, structural phase, and subsequently superconducting properties after the growth was completed and while the film was cooling down.

To test this hypothesis, we performed growths of three NbN films with identical conditions: $T_s= 1000~\degree$C and 400 W power for the nitrogen plasma. For the first growth, we turned off the plasma power and nitrogen gas flow immediately after the NbN growth was finished, as had been done for all growths up to this point in this report. For the second growth, we left the nitrogen gas flow and plasma power on after the NbN growth was finished and while the sample was cooling down, though the shutter for the nitrogen source was closed. We hypothesized that maintaining the presence of an active nitrogen flux in the chamber after growth would both prevent loss of nitrogen after growth and potentially lead to additional absorption of nitrogen by the NbN while cooling from the growth temperature. In the third growth, an AlN film approximately 20 nm in thickness was grown on the NbN immediately upon completion of the NbN growth. We hypothesized that, given that the nitrogen diffusion coefficient in AlN is very small at these temperatures \cite{Klomp1990ALUMINIUMREACTIONS}, the AlN film would prevent any changes in the nitrogen composition in the NbN after growth by acting as a diffusion barrier.

We assessed the results of these growths using XRD and R vs T measurements. Figure \ref{fig:Plasma_post_growth}(e) shows the symmetric NbN peak for all 3 films. We observe clear variation in the NbN lattice parameter, with the "plasma it left on" and "capped with AlN" films showing larger lattice parameters, closer to that of the higher nitrogen composition $\delta$ phase, than that of the "plasma turned off" film, which is closer to that of the lower nitrogen composition $\gamma$ phase. All films exhibit evidence of the $\gamma$  structure in RSM measurements, though due to similar structures and nearby peaks, we cannot conclusively discern if any of the films are mixed $\delta$ and $\gamma$ structure.

R vs T measurements (Figure \ref{fig:Plasma_post_growth}(d)) show corresponding changes in the superconducting properties of each film. Both the "plasma left on" and "capped with AlN" films indicate a transition to the zero resistance state at around 12 K, well above the transition of the "plasma turned off" film, which transistions at around 7 K. We attribute the enhanced superconductivity of the "capped with AlN" and "plasma left on" films to the effectiveness of both treatments to prevent the loss of nitrogen from the films after growth, which leads to structural changes that reduce the superconducting critical temperatures of the films. Given the onset of superconducting fluctuations nearer to 17 K for the "plasma left on" film, the presence of some $\delta$ phase material in the film seems likely.

\section{Discussion}\label{sec:discussion}

There are several implications of this study regarding the growth of NbN thin films by MBE, as well as by other growth methods. We note our lack of observation of pure $\delta$ phase material possessing a superconducting critical temperature near 17 K, the maximum reported critical temperature for $\delta$-NbN reported in  studies by Pan et al. and Keskar et al., both of which use sputter deposition and substrate temepratures of approximately 700 \degree C\cite{Pan1983InvestigationFilms,Keskar1971}. Several theoretical calculations have predicted\cite{Isaev2007PhononStudy,Ivashchenko2010PhaseNitrides}, and several experimental studies have demonstrated \cite{Brauer1960NitridesNiobium,Lengauer2000Phase1400C}, the instability of B1 structure $\delta$-NbN at room temperature. HTXRD studies demonstrate that the $\delta$ phase appears as a stable phase only at temperatures above approximately $1100~\degree$C \cite{Lengauer2000Phase1400C}. Below this temperature it converts either to $\gamma$ or to $\epsilon$, depending on the composition of the film.

A previous study by Lengauer et al. reported that the maximum transition temperature of NbN is observed for films that approach a 1:1 N:Nb composition \cite{Lengauer1990a}. Several studies have also discussed the observation that achieving the maximum critical temperature of NbN appears limited by the instability of stoichiometric $\delta$-NbN. Oya stated that, “the maximum $T_c$ of $\delta$-NbN seems to be limited by the $\delta$-NbN/$\epsilon$-NbN transformation\cite{Oya1974}.” Wolf et al. reported that the maximum $T_c$ for sputtered NbN samples occurred for samples which exhibited the coincidence of $\delta$ and $\epsilon$ phase\cite{Wolf1979EffectsFilms}. Haase reported that, while NbN films of pure $\delta$ phase could be grown by sputter deposition with a $T_c$ up to 15.9 K, films which exhibited a higher $T_c$ of 16.7 K all exhibited large fractions of $\epsilon$ phase material, another example of the $T_c$ of NbN being limited by the $\delta$ to $\epsilon$ transition\cite{Haase1988ImprovementsSputtering}. This is true for our study as well, with the maximum $T_c=15.0$ K occurring for a sample which exhibits the coincidence of $\delta$ and $\epsilon$ phase.

We believe that our findings support Oya's statement that the transition to the $\epsilon$ phase can prohibit the observation of the maximum $T_c$ of NbN. As can be seen in Figure \ref{fig:R_vs_T}(c), the critical temperature is increasing as the substrate temperature is decreasing from $T_s=1050~\degree$C to $T_s=850~\degree$C. The critical temperature is also increasing as $T_s$ increases from $T_s=650~\degree$C to $T_s=700~\degree$C. Given our finding that substrate temperature effectively controls film composition, previous results that the composition window of $\epsilon$-NbN is very narrow and is nearly stoichiometric, and the results that show that the maximum $T_c$ of NbN is observed for stoichiometric $\delta$-NbN, we conclude that the maximum $T_c$ of our films is limited by the transition of the films to the $\epsilon$-NbN phase when the composition approaches 50\% nitrogen, which for our growth conditions occurs at a substrate temperature between $800~\degree \textrm{C} < T_s < 850~\degree$C.

There are potentially several routes to stabilizing the $\delta$ phase to achieve phase pure NbN films by plasma MBE with higher $T_c$ than the maximum $T_c\cong15$ K we report here. Musenich et al. suggest that the high carbon content of many sputtered NbN films serves to stabilize the $\delta$ phase\cite{Musenich1994GrowthTemperature}, which is plausible given the complete substitution of carbon and nitrogen within the $\delta$ phase of $\textrm{NbN}_{1-x}C_x$\cite{Brauer1960NitridesNiobium}, and the lack of a hexagonal $\epsilon$-NbC phase\cite{Toth1971}. Wolf reported carbon content of several percent for sputtered NbN films, though the source of the carbon was not identified\cite{Wolf1979EffectsFilms}. SIMS analysis reveals that the carbon content of our films is approximately 0.2\%.

Interestingly, Haase demonstrated that tuning the deposition parameters of NbN to maximize $T_c$ led to mixed $\epsilon$/$\delta$ phase films with $T_c=16.7~K$, but through co-deposition of Al during NbN growth, pure $\delta$-NbN  films with $T_c$ up to 16.9 K could be produced. The Al composition of the films was below the detection limit of their measurement of $0.03 \%$ atomic percent, and films which possessed detectable Al compositions of approximately 1\% all showed reduced $T_c$. Haase hypothesized that it was actually not the incorporation of the Al into the crystal, but the effect of the Al on the film growth that led to the change, though this hypothesis was not substantiated. We suggest that, given the promising result from Haase and the precise control of Al flux available from an effusion cell, further work of NbN growth by MBE should explore the doping of NbN with Al with the goal of enhancing the stability of the $\delta$ phase at nitrogen compositions which yield films with the highest critical temperatures.

Briefly, we would like to comment on the transition from $\delta$ to $\gamma$ phase as the substrate temperature is increased. Although there is a history of extensive discussion of the $\gamma$ to $\delta$ transition, the distinction of these phases experimentally in epitaxial thin-film form is somewhat subtle. We suggest that the appearance of double peaks in an asymmetric XRD measurement in the location where a single peak is expected for cubic $\delta$-NbN is strongly suggestive of $\gamma$ phase material. Several recent reports on NbN thin film growth present asymmetric XRD evidence of double peaks in locations expected for epitaxial $\gamma$-$\textrm{Nb}_4\textrm{N}_3$, as well as critical temperatures that match the expected value for $\gamma$-$\textrm{Nb}_4\textrm{N}_3$ \cite{Nepal2016CharacterizationSubstrates}\cite{Kobayashi2020CoherentSputtering}, but in these studies they propose alternative explanations for the double peaks, and do not discuss their consistency with the $\gamma$ structure.

We believe that we present in this study the first measurement of the superconducting transition in phase pure $\beta$-$\textrm{Nb}_2\textrm{N}$. With a critical temperature below 1 K, we conclude that superconducting $\beta$-$\textrm{Nb}_2\textrm{N}$ devices are likely limited to dilution refrigerator environments. Nevertheless, as has been discussed previously, $\beta$-$\textrm{Nb}_2\textrm{N}$ is an exciting material from the perspective of nitride heteroepitaxy. $\beta$-$\textrm{Nb}_2\textrm{N}$ has a close-packed hexagonal cation sublattice which is nearly isostructural with that of both AlN and GaN. Thereby our results on the growth of $\beta$-$\textrm{Nb}_2\textrm{N}$ thin films create the opportunity to integrate with the III-N semiconductors films which are highly crystalline, atomicly flat, metallic and superconducting, and which possess a high degree of structural similarity to III-N semiconductors. This can potentially enable unprecedented levels of crystalline perfection in devices such as as Josephson junctions, acoustic resonators, and in general, multilayers and superlattices of semiconductor and metal, and semiconductor and superconductor layers. 

\section{Methods}\label{sec:methods}

Growth of the NbN films was performed in a Veeco GENxplor MBE system. Nb of 99.95\% purity (excluding Ta) is supplied by an electron-beam evaporator; Nb flux is measured using an electron impact energy spectroscopy (EIES) system.  Active nitrogen is supplied using a RF plasma source. For 200 W nitrogen RF plasma power, we use a nitrogen flow rate of 1.95 SCCM. For 400 W nitrogen RF plasma power, we use a nitrogen flow rate of 3 SCCM. For 500 W nitrogen RF plasma power, we use a nitrogen flow rate of 4.5 SCCM.
Film growth is monitored in situ using reflection high energy electron diffraction (RHEED) with an accelerating voltage of 15 kV and a current of 1.5 A.

c-plane sapphire pieces $1~\textrm{cm}\times1~\textrm{cm}$ are mounted in solid molybdenum holders during growth. Substrate temperature is monitored using a thermocouple behind the sample, and we expect the substrate thermocouple does not accurately measure the true surface temperature of the growing film\cite{ScottKatzer2020}. The difficulty in accurately measuring the surface film temperature is further compounded by the fact that the growth of the metallic NbN film itself significantly alters the substrate temperature due to increased absorption of thermal radiaiton from the substrate heaters. Based on RHEED measurements of the desorption of accumulated Ga deposited on the surface of NbN, we estimate that the surface temperature after growth of 50 nm of NbN is approximately $100~\degree$C higher than the reading from the substrate thermocouple\cite{Heying2000ControlEpitaxy}.

The surface properties of the NbN films were characterized using tapping mode atomic force microscopy (AFM) using an Asylum Research Cypher AFM. The superconducting transition temperatures of NbN films were measured by performing $R$ vs $T$ measurements in a Quantum Design PPMS system, and $T_c$ is defined as the temperature at which the film resistance goes to half of the normal state resistance. The transition width is defined as the temperature difference between the temperature at which the resistance reaches 90\% of the normal state resistance and the temperature at which the resistance reaches 10\% of the normal state resistance.

High-resolution X-ray diffraction (XRD) coupled scans were performed using a Malvern Panalytical X'pert Pro system with a with Cu K$\alpha$1 radiation (1.54057 $\si{\angstrom}$) X-ray source operated at 45 kV, 40 mA. XRD coupled scans measurements were performed using a triple axis geometry. XRD-RSM measurements were performed using a Malvern Panalytical Empyrean diffractometer operated with a Cu K$\alpha$1 radiation (1.54057 $\si{\angstrom}$) X-ray source operated at 45 kV, 40 mA, with a pixel detector operated in a 1D mode. EDS measurements were performed in a Tescan Mira3 field emission SEM. Measurements were performed with a 5 keV accelerating voltage. The measurements were made as line scans across a length of $200~\mu m$. The quantitative EDS analysis was performed using a standardless P/B-ZAF algorithm

\medskip

\medskip
\textbf{Acknowledgements} \par 
This work used the CNF, CCMR and CESI Shared Facilities partly supported by the NSF NNCI program (NNCI-2025233), MRSEC program (DMR-1719875) and MRI DMR-1338010 and Kavli Institute at Cornell (KIC). We acknowledge funding support from the Office of Naval Research, monitored by P. Maki under award numbers N00014-20-1-2176 and N00014-17-1-2414.

\medskip

%
\bibliographystyle{MSP}
\bibliography{references}




\end{document}